\documentclass[aps,preprint,showpacs]{revtex4}%
\usepackage{amsfonts}
\usepackage{amsmath}
\usepackage{amssymb}
\usepackage{graphicx}%
\begin{document}
\title[ ]{Auxiliary fields in open gauge theories}
\author{N. Djeghloul, A. Meziane, and M. Tahiri}
\affiliation{Laboratoire de Physique Th\'{e}orique d'Oran (LPTO), Universit\'{e} d'Oran, BP
1524 El M'Naouer, 31100 Es-Senia, Oran, Algeria.}


\begin{abstract}
We show that for open gauge theories, it is possible to build an off-shell
Becchi--Rouet--Stora--Tyutin (BRST) algebra together with an invariant
extension of the classical action. This is based on the introduction of
auxiliary fields, after having defined an on-shell invariant quantum action,
where the gauge-fixing action is written as in Yang--Mills type theories up to
a modified BRST operator. An application to simple supergravity is performed.


\end{abstract}
\startpage{01}
\endpage{ }
\maketitle

\section{\bigskip Introduction}

The Batalin--Vilkovisky (BV) formalism \cite{batalin} currently appears to be
the most powerful method for quantizing general gauge theories, i.e. gauge
theories which are reducible and/or with an open algebra (for a review, see
Ref. \cite{Hen}). It leads to the construction of the quantum theory in which
effective BRST transformations are nilpotent on shell. This is realized by
doubling the total number of gauge fields and ghost fields by introducing
corresponding antifields, which are then eliminated by means of a gauge
fermion functional containing the gauge-fixing conditions associated to all
the invariances of the classical action defining the gauge theory.

On the other hand, it is well known that in supersymmetric theories, in
particular for supergravity theories, which represent prototypes of open gauge
theories, the superspace formalism enables ones to close the gauge algebra
through the introduction of auxiliary fields and then gives the possibility to
quantize such theories by using the standard BRST formalism (for a review, see
Ref. \cite{Van-Nieuwen}). This is also the case of topological antisymmetric
tensor gauge theories, so-called BF theories, which represent prototypes of
reducible gauge theories, where auxiliary fields can be introduced in terms of
a BRST superspace \cite{tahiri1}. Let us note that the same geometrical
approach has been considered in order to give another possibility leading to
the standard minimal set of auxiliary fields in simple supergravity
\cite{tahiri2}. In an alternative way, it was shown that an off-shell
formulation of simple supergravity in terms of a principal superfiber bundle
can also be performed \cite{tahiri3}. Moreover, still in the context of simple
supergravity, the introduction of auxiliary fields can be realized via the BV
formalism \cite{anto1} (see also Ref. \cite{anto2} for the case of reducible
gauge theories).

Furthermore, we have shown for BF theories how the structure of auxiliary
fields with nonvanishing ghost numbers as well as the invariant extension of
the classical action come from an on-shell BRST invariant quantum action
\cite{tahiri4}. The latter was simply constructed by writing the gauge-fixing
action as in Yang--Mills theories by modifying the classical BRST operator.

It would be worthwhile to extend the analysis developed in Ref. \cite{tahiri4}
in order to discuss general gauge theories, independent of the underlying
classical action. In this paper we will be interested to build an on-shell
invariant gauge-fixed action for irreducible open gauge theories which permits
us to introduce auxiliary fields and to determine an invariant extension of
the classical action in such theories. The obtained auxiliary fields are of
ghost number $(-1)$ and thus cannot be considered classical fields. But even
if these fields are objectively, at least at a formal level, different from
those that appear in known open algebra theories, they have the crucial
advantage that they can be introduced via a systematic procedure applied to
any given gauge theory plagued with a symmetry algebra that closes only on
shell. It is worth noting that a formal approach for introducing standard
auxiliary fields (of ghost number zero) was also proposed in Ref.
\cite{tahiri5}.

As an application we focus on the case of simple supergravity for which the
introduced method to build up an off-shell BRST operator leads to a set of
nonclassical auxiliary bosonic spinor fields of ghost number -1.

\section{\bigskip\label{Sec2}On shell formulation}

Let us start from a gauge system described by a classical action $S^{cl}$
depending on the gauge fields ${\phi}^{i}$ with parity $\epsilon_{i}$. The
invariance of $S^{cl}$ under gauge transformations, $\delta\phi^{i}=R_{\alpha
}^{i}\varepsilon^{\alpha}$, leads to the Noether identities, $S_{,i}%
^{cl}R_{\alpha}^{i}=0$, where $X,_{i}$ denotes the variation of $X$ with
respect to $\phi^{i}$. The generators $R_{\alpha}^{i}$ are operators acting on
the gauge parameters $\varepsilon^{\alpha}$ with parity $\epsilon_{\alpha}$.
The gauge functions and their associated gauge equations which characterize
the gauge algebra depend on the nature of the gauge theory \cite{batalin,Hen}.
In the following we restrict ourselves to irreducible open gauge theories,
i.e. all generators $R_{\alpha}^{i}$ are independent and the commutator of two
gauge transformations leads to the definition of two gauge functions
$G_{\beta\gamma}^{\alpha}$ and $G_{\alpha\beta}^{ij}$, satisfying
$R_{\alpha,j}^{i}R_{\beta}^{j}-(-1)^{\epsilon_{\alpha}\epsilon_{\beta}%
}R_{\beta,j}^{i}R_{\alpha}^{j}=R_{\gamma}^{i}G_{\alpha\beta}^{\gamma}%
+S_{,j}^{cl}G_{\alpha\beta}^{ij}$, which means that the commutator is closed
up to equations of motion. Upon introducing the ghost $c^{\alpha}$ with parity
$(\epsilon_{\alpha}+1)$, the above commutator can be put in the following form%

\begin{equation}
G^{i}_{,j}G^{j} + G^{i}_{,\alpha}G^{\alpha} + S^{cl}_{,j}G^{ij} = 0,
\label{phi}%
\end{equation}
where $G^{i}= R^{i}_{\alpha}c^{\alpha}$, $G^{\alpha} = \frac{1}{2}(
-1)^{\epsilon_{\beta}} G^{\alpha}_{\beta\gamma}c^{\gamma}c^{\beta}$, $G^{ij}=
\frac{1}{2}( -1)^{\epsilon_{\alpha}} G^{ij}_{\alpha\beta}c^{\beta}c^{\alpha}$,
and $X_{,\alpha}$ denotes the variation of $X$ with respect to $c^{\alpha}$.

The classical BRST transformations of the fields $\phi^{i}$ are simply
obtained as usual by replacing $\varepsilon^{\alpha}$ by $c^{\alpha}$, we have%

\begin{equation}
Q\phi^{i}=G^{i}. \label{phi1}%
\end{equation}
In view of Eq. (\ref{phi}), the on-shell nilpotency of $Q$ acting on $\phi
^{i}$, i.e. $Q^{2}\phi^{i}=-S_{,j}^{cl}G^{ij}$, is ensured provided that
\begin{equation}
Qc^{\alpha}=G^{\alpha}. \label{phi2}%
\end{equation}
Furthermore, to express the on-shell nilpotency of $Q$ acting on $c^{\alpha}$,
a new gauge function $G^{\alpha i}$ is also needed, in order to write
$Q^{2}c^{\alpha}=-S_{,i}^{cl}G^{\alpha i}$ , and according to Eq.
(\ref{phi2}), we obtain the following gauge equation:
\begin{equation}
G_{,i}^{\alpha}G^{i}+G_{,\beta}^{\alpha}G^{\beta}+S_{,i}^{cl}G^{\alpha i}=0.
\label{phi3}%
\end{equation}
It is the Jacobi identity which leads to the definition of $G^{\alpha i}$ as
well as of a second new gauge function $G^{ijk}$. Besides $G^{\alpha}$,
$G^{ij}$ $(G^{\alpha i}$, $G^{ijk})$ which are quadratic (cubic) in
$c^{\alpha}$, it is possible to introduce other gauge functions $G^{ijkl}$,
$G^{\alpha ij},...,$ which are higher-order polynomials in $c^{\alpha}$, by
using higher-order commutators of the generators $R_{\alpha}^{i}$ \cite{Hen}.
Let us note that in the realm of the BV formalism, the gauge algebra is also
generated by the classical master equation \cite{batalin,Hen}.

We shall mention that the known open gauge theories (e.g., supergravity
theories) are described by a gauge algebra in which the set of gauge functions
contains only $G^{\alpha}$, $G^{ij}$, and $G^{\alpha i}$, and all the
remaining gauge functions $G^{ijk}$, $G^{ijkl}$ $G^{\alpha ij}$,..., vanish.
Thus, for simplicity and to present computations leading to insight in the
generalization of the analysis in Ref. \cite{tahiri4} (see also Ref.
\cite{tahiri3}) to open gauge theories, we consider an open gauge algebra that
is characterized by the three nonvanishing gauge functions $G^{\alpha}$,
$G^{ij}$, and $G^{\alpha i}$. In addition to Eqs. (\ref{phi}) and
(\ref{phi3}), new identities need to be satisfied. The latter follow from the
higher-order gauge equations \cite{batalin, Hen}, in which we take off the
vanishing gauge functions; we find the following nontrivial identities:%

\begin{equation}
G_{,k}^{ij}G^{k}+G_{,\alpha}^{ij}G^{\alpha}-(-1)^{\epsilon_{j}}\{G_{,k}%
^{i}G^{kj}+G_{,\alpha}^{i}G^{\alpha j}\}+(-1)^{\epsilon_{j}(\epsilon_{i}%
+1)}\{{G_{,k}^{j}G^{ki}+G_{,\alpha}^{j}G^{\alpha i}}\}=0, \label{BRST1}%
\end{equation}%
\begin{equation}
G_{,l}^{ij}G^{lk}+G_{,\alpha}^{ij}G^{\alpha k}+(-1)^{\epsilon_{k}(\epsilon
_{i}+\epsilon_{j})}\{G_{,l}^{jk}G^{li}+G_{,\alpha}^{jk}G^{\alpha
i}\}+(-1)^{\epsilon_{j}(\epsilon_{i}+\epsilon_{k})}\{{G_{,l}^{ki}%
G^{lj}+G_{,\alpha}^{ki}G^{\alpha j}}\}=0. \label{BRST2}%
\end{equation}
In what follows we turn to discuss how to construct the quantum theory of an
open gauge theory characterized by a classical BRST algebra given by Eqs.
(\ref{phi})--(\ref{BRST2}). It is obvious that a $Q$-exact form of the
gauge-fixing action cannot be suitable to build the full invariant quantum
action because of the on-shell nilpotency of the classical BRST operator $Q$.
To this end, we generalize the prescription discussed in Ref. \cite{tahiri4}
by simply modifying the classical BRST operator $Q$. Accordingly, the
gauge-fixing action written as in Yang--Mills type theories must also be
modified so that the complete quantum action becomes invariant on shell. We
first introduce a gauge fermion $\psi$ to implement gauge constraints,
$F^{\alpha}=0$, associated to all the invariances of the classical action
$S^{cl}$; we have
\begin{equation}
\psi=\bar{c}_{\alpha}F^{\alpha}, \label{BRST 3}%
\end{equation}
where $\bar{c}_{\alpha}$ represent the antighosts, which allow us as usual to
define the Stueckelberg auxiliary fields $b_{\alpha}$ through the action of
$Q$, so that%

\begin{equation}
Q\bar{c}_{\alpha}=b_{\alpha}, \quad\quad Q b_{\alpha}=0. \label{BRST4}%
\end{equation}
Let us note that the gauge-fixing functions $F^{\alpha}$ may depend only on
the gauge fields $\phi^{i}$, since the gauge symmetries are considered to be irreducible.

So, at the quantum level we define a modified BRST operator $\Delta$,%

\begin{equation}
\Delta=Q+\tilde{Q}, \label{BRST55}%
\end{equation}
satisfying $\Delta^{2}=0$, up to equations of motion, and $\Delta S^{q}=0$,
where $S^{q}$ is the quantum action. As discussed above, the gauge-fixing
action $S^{gf}$ in $S^{q}$ cannot be cast in the form $S^{gf}=Q\psi$. To
rectify this we modify $S^{gf}$ so that $Q$ is replaced by $(Q+x\tilde{Q})$;
i.e., we have%

\begin{equation}
S^{q}= S^{cl}+ (Q + x\tilde{Q})\psi. \label{BRST66}%
\end{equation}
We remark that $\tilde{Q}$ has vanishing action on the pairs $(\bar{c}%
_{\alpha}, b_{\alpha})$. This simply follows from Eq. (\ref{BRST4}) which says
that the nilpotency on those fields is already guaranteed. To derive the
action of $\tilde{Q}$ on the fields $\phi^{i}$ and $c^{\alpha}$, we use the
structure of the open gauge algebra together with the invariance of the
quantum action $S^{q}$, as written in Eq. (\ref{BRST66}), under the on-shell
nilpotent quantum BRST operator, as defined in Eq. (\ref{BRST55}).

In view of the on-shell nilpotency of the classical BRST operator $Q$, the
variation of $S^{q}$ under the quantum BRST $\Delta$ can be written as%

\begin{equation}
\Delta S^{q} = S^{cl}_{,i}(\tilde{Q}\phi^{i} - (-1)^{\epsilon_{i}(\epsilon
_{j}+1)}\psi_{,j}G^{ji}) + (\tilde{Q}Q + xQ\tilde{Q} + x\tilde{Q}^{2})\psi.
\label{BRST68}%
\end{equation}
To guarantee the invariance of $S^{q}$ under $\Delta$, we note that by
choosing
\begin{equation}
\tilde{Q}\phi^{i}= - (-1)^{\epsilon_{i}}\psi_{,j}G^{ij} \label{BRST69}%
\end{equation}
the first term on the right-hand side of Eq. (\ref{BRST68}) vanishes.
Substituting Eq. (\ref{BRST69}) into Eq. (\ref{BRST68}) and using the identity
given by Eq. (\ref{BRST1}), we get
\begin{equation}
\Delta S^{q} = \psi_{,j} R^{j}_{\alpha} \{\tilde{Q}c^{\alpha} -
(-1)^{(\epsilon_{\alpha}+1)(\epsilon_{i}+1)}\psi_{,i}G^{\alpha i}\}\nonumber\\
\end{equation}
\begin{equation}
+ (-1)^{\epsilon_{j}}x\psi_{,j}\psi_{,i} \{(-1)^{\epsilon_{k}+1} G^{ji}%
_{,k}\psi_{,l}G^{lk} + G^{ji}_{,\alpha}\tilde{Q}c^{\alpha}\}\nonumber\\
\end{equation}
\begin{equation}
+ (-1)^{\epsilon_{i}}(1-2x)\{ (-1)^{\epsilon_{\alpha}}b_{\alpha}F^{\alpha
}_{,j}\psi_{,i}G^{ji} + (-1)^{\epsilon_{k}} \psi_{,i}\psi_{,j}\psi_{,k}
G^{jk}G^{i}\}\nonumber\\
\end{equation}
\begin{equation}
+ (-1)^{\epsilon_{j}}(x- \frac{1}{2})\psi_{,j}\psi_{,i} \{ G^{ji}_{,k} G^{k} +
G^{ji}_{,\alpha}G^{\alpha} \}. \label{gaugefixing}%
\end{equation}
Further we learn from Eq. (\ref{gaugefixing}) that by taking
\begin{equation}
\tilde{Q}c^{\alpha}= (-1)^{(\epsilon_{\alpha}+1) (\epsilon_{i}+1)}\psi
_{,i}G^{\alpha i}, \label{BRST70}%
\end{equation}
\begin{equation}
x = \frac{1}{2}, \label{BRST71}%
\end{equation}
the $\Delta$-invariance of $S^{q}$ is completely ensured. We remark that the
vanishing of the second term on the right-hand side of Eq. (\ref{gaugefixing})
follows from the use of the identity given by Eq. (\ref{BRST2}).

Thus, we have obtained the full quantum action $S^{q}$,
\begin{equation}
S^{q}=S^{cl}+\frac{1}{2}(-1)^{\epsilon_{i}}\psi_{,i}\psi_{,j}G^{ij}+\Delta
\psi, \label{BRST72}%
\end{equation}
invariant under the BRST operator $\Delta$ determined by Eqs. (\ref{BRST55}),
(\ref{BRST69}), and (\ref{BRST70}) together with Eqs. (\ref{phi1}),
(\ref{phi2}), and (\ref{BRST4}), which is nilpotent on shell. In fact, after a
similar straightforward computation, we get
\begin{equation}
\Delta^{2}\phi^{i}=-S_{,j}^{q}G^{ij}-S_{,\alpha}^{q}G^{\alpha i},
\end{equation}
\label{BRST73}%
\begin{equation}
\Delta^{2}c^{\alpha}=-S_{,i}^{q}G^{\alpha i}. \label{BRST74}%
\end{equation}
\ It is remarkable that the used prescription, which simply consists in the
modification of the classical BRST operator and of the gauge-fixing action
written as in Yang--Mills theories, provides an on-shell quantization, where
in particular the quantum action contains four-ghost couplings. The latter are
characteristic for open gauge theories like supergravity theories
\cite{Van-Nieuwen}.

\section{\bigskip\label{Sec3}Auxiliary fields}

Let us now discuss how we can introduce auxiliary fields, as a generalization
of the approach developed in Ref. \cite{tahiri4}, so that we end up with an
off-shell structure for open gauge theories. For this purpose, we start with
the following BRST transformations
\begin{equation}
\Delta\phi^{i}=G^{i}-(-1)^{\epsilon_{i}\epsilon_{j}}G^{ij}\eta_{j},
\label{BRST75}%
\end{equation}%
\begin{equation}
\Delta c^{\alpha}=G^{\alpha}+G^{\alpha i}\eta_{i}, \label{BRST76}%
\end{equation}%
\begin{equation}
\Delta\bar{c}_{\alpha}=b_{\alpha},\quad\quad\quad\Delta b_{\alpha}=0.
\label{BRST77}%
\end{equation}
These follow from those which are nilpotent on shell by replacing $\psi_{,j}$
by $\eta_{j}$. Making the same replacement in Eq. (\ref{BRST72}), we put the
quantum action $S^{q}$ in the form
\begin{equation}
S^{q}=S^{cl}+\frac{1}{2}(-1)^{\epsilon_{j}}G^{ij}\eta_{i}\eta_{j}+\Delta\psi.
\label{BRST78}%
\end{equation}
At this point, by assuming that the $\eta_{i}$ are now true fields of parity
$(\epsilon_{i}+1)$ and ghost number $(-1)$, it is worth noting that the
quantum action $S^{q}$ allows us to see that they are auxiliary, nondynamical
fields as their equations of motion are constraints,
\begin{equation}
(-1)^{\epsilon_{j}(\epsilon_{i}+1)}G^{ij}(\psi_{,j}-\eta_{j})=0.
\label{BRST79}%
\end{equation}
Indeed, the only terms of the quantum action contributing to the equations of
motion of the fields $\eta_{i}$ are $(\frac{1}{2}(-1)^{\epsilon_{j}%
(\epsilon_{i}+1)}G^{ij}\eta_{i}\eta_{j})$ and $(-(-1)^{\epsilon_{j}%
(\epsilon_{i}+1)}G^{ij}\psi_{,i}\eta_{j})$. The last term follows from the
gauge-fixing action $\Delta\psi$ by using the transformations of Eq.
(\ref{BRST75}). However, substituting Eq. (\ref{BRST79}) into Eqs.
(\ref{BRST75})-(\ref{BRST78}), which is equivalent to replace $\eta_{i}$ by
$\psi_{,i}$, again we obtain the quantum action and its on-shell BRST symmetry.

Further, we shall determine the action of the BRST operator on these auxiliary
fields, so that the BRST algebra closes off shell. This is simply realized by
imposing the off-shell nilpotency condition $\Delta^{2}=0$. So, we obtain%

\begin{equation}
\Delta\eta_{i}=-(-1)^{\epsilon_{j}(\epsilon_{i}+1)}G_{,i}^{j}\eta
_{j}-(-1)^{\epsilon_{i}}S_{,i}^{cl}. \label{BRST80}%
\end{equation}
It is easy to check the off-shell nilpotency of Eqs. (\ref{BRST75}) and
(\ref{BRST76}) by an explicit calculation. We note, in particular , that in
deriving Eq. (\ref{BRST80}) we have used the identities given by Eqs.
(\ref{BRST1}) and (\ref{BRST2}), which can be cast in the following form
\begin{equation}
(-1)^{\epsilon_{j}(\epsilon_{i+1})}\eta_{j}\eta_{i}(\frac{1}{2}G_{,k}%
^{ij}G^{k}+\frac{1}{2}G_{,\alpha}^{ij}G^{\alpha}-(-1)^{\epsilon_{j}}G_{,k}%
^{i}G^{kj}-(-1)^{\epsilon_{j}}G_{,\alpha}^{i}G^{\alpha j})=0, \label{BRST81}%
\end{equation}%
\begin{equation}
(-1)^{\epsilon_{j}(\epsilon_{i+1})}\eta_{k}\eta_{j}\eta_{i}(\frac{1}{2}%
G_{,l}^{ij}G^{lk}+\frac{1}{2}G_{,\alpha}^{ij}G^{\alpha k})=0. \label{BRST82}%
\end{equation}
Moreover, after a similar straightforward calculation, we find that
\begin{equation}
S_{inv}=S^{cl}+\frac{1}{2}(-1)^{\epsilon_{j}}G^{ij}\eta_{i}\eta_{j}.
\label{Sinv}%
\end{equation}
\label{BRST67}represents the $\Delta$-invariant extension of the classical
action $S^{cl}$.

\section{Case of D=4, N=1 Supergravity}

The physical fields content of simple supergravity \cite{Van-Nieuwen} is given
by the vierbein $e_{\mu}^{a}$, and the gravitino $\psi_{\mu}^{A}$ with
$a=1,...,4$ label the flat Minkowski space, $\mu=1,...,4$ labels the curved
Riemannian space and $A=1,...,4$ is related to the $N=1$ supersymmetry. The
classical action of the model is given by%

\begin{equation}
S^{cl}=\frac{1}{2}ee_{a}^{\mu}e_{b}^{\nu}R_{\mu\nu}^{ab}-\frac{1}%
{4}\varepsilon^{\mu\nu\rho\sigma}\bar{\psi}_{\mu}\gamma_{5}\gamma_{\nu}%
S_{\rho\sigma}, \label{SG Action}%
\end{equation}
where $e=\det(e_{\mu}^{a})$, $R_{\mu\nu}^{ab}=$ $\partial_{\mu}\omega_{\nu
}^{ab}+\frac{1}{2}(\omega_{\mu}^{ad}\omega_{\nu d}^{b}-\omega_{\mu}^{bd}%
\omega_{\nu d}^{a})-(\mu\leftrightarrow\nu)$ is the Lorentz curvature,
$\gamma_{5}=i\gamma_{0}\gamma_{1}\gamma_{2}\gamma_{3}$, $S_{\rho\sigma
}=\partial_{\rho}\psi_{\sigma}+\frac{1}{2}\omega_{\rho}^{ab}\sigma_{ab}%
\psi_{\sigma}-(\rho\leftrightarrow\sigma)$ with $\sigma_{ab}=\frac{1}%
{4}[\gamma_{a},\gamma_{b}]$ is the Fermi curvature, $\bar{\psi}_{\nu}%
=\psi_{\nu}^{T}C$, $C$ \ is the charge conjugation matrix, $\gamma_{\nu
}=e_{\nu}^{a}\gamma_{a}$, $\gamma_{a}$ are the Dirac matrices. One recalls
that the theory admits a vanishing torsion leading to a nonpropagating spin
connection $\omega_{\mu}^{ab}$, which therefore can be expressed in terms of
$e_{\mu}^{a}$ and $\psi_{\mu}^{A}$, i.e., $\omega_{\mu}^{ab}=\frac{1}{2}e^{\nu
b}[\{\partial_{\nu}e_{\mu}^{a}-\frac{1}{4}\bar{\psi}_{\nu}\gamma^{a}\psi_{\mu
}-(\mu\leftrightarrow\nu)\}+e_{\mu}^{d}e^{\rho a}\{\partial_{\nu}e_{\rho
a}-\frac{1}{4}\bar{\psi}_{\nu}\gamma_{d}\psi_{\rho}-(\nu\leftrightarrow
\rho)\}]$; $e_{a}^{\mu}$ is the inverse of the vierbein defined by $e_{a}%
^{\mu}e_{\nu}^{a}=\delta_{\nu}^{\mu}$ and $e_{a}^{\mu}e_{\mu}^{b}=\delta
_{a}^{b}$.

The symmetries of the model are diffeomorphism (general coordinates
transformations), Lorentz rotations and supersymmetry. So the action
(\ref{SG Action}) is invariant under the transformation \cite{Van-Nieuwen}
(expressed \textit{\`{a} la} BRST)%
\begin{align}
Qe_{\mu}^{a}  &  =c^{\nu}\partial_{\nu}e_{\mu}^{a}+\partial_{\mu}c^{\nu}%
e_{\nu}^{a}-c^{ab}e_{\mu b}+\frac{1}{2}\bar{c}\gamma^{a}\psi_{\mu},\nonumber\\
Q\psi_{\mu}  &  =c^{\nu}\partial_{\nu}\psi_{\mu}+\partial_{\mu}c^{\nu}%
\psi_{\nu}-\frac{1}{2}c^{ab}\sigma_{ab}\psi_{\mu}+\partial_{\mu}c+\frac{1}%
{2}\omega_{\mu}^{ab}\sigma_{ab}c, \label{SG symm}%
\end{align}
where $c^{\nu}$, $c^{ab}$, and $c$ are, respectively, the ghost fields
associated to local diffeomorphism, Lorentz rotations, and supersymmetry
parameters. These ghost fields are all of ghost number +1; $c^{\nu}$and
$c^{ab}$ are fermionic while $c$ is bosonic. One can complete the action of
the BRST operator (\ref{SG symm}) on the ghosts fields by%
\begin{align}
Qc^{\mu}  &  =c^{\nu}\partial_{\nu}c^{\mu}-\frac{1}{4}\bar{c}\gamma^{\mu
}c,\nonumber\\
Qc^{ab}  &  =c^{\nu}\partial_{\nu}c^{ab}-c^{ad}c_{d}^{b}+\frac{1}{4}\bar
{c}\gamma^{\mu}c\omega_{\mu}^{ab},\nonumber\\
Qc  &  =c^{\nu}\partial_{\nu}c-\frac{1}{2}c^{ab}\sigma_{ab}c+\frac{1}{4}%
\bar{c}\gamma^{\mu}c\psi_{\mu}, \label{SG ghost symm}%
\end{align}
leading to the following on-shell property of the BRST operator $Q$:
\begin{align}
Q^{2}\psi_{\mu}  &  =-G_{\mu\nu}\frac{\delta S_{0}}{\delta\bar{\psi}_{\nu}%
},\label{on shell supergravity}\\
Q^{2}c^{ab}  &  =-G_{\mu}^{ab}\frac{\delta S_{0}}{\delta\bar{\psi}_{\mu}},\\
Q^{2}X  &  =0\text{ \quad for all others fields,}%
\end{align}
where the equation of motion of the gravitino reads%
\begin{equation}
\frac{\delta S_{0}}{\delta\bar{\psi}_{\mu}}=-\frac{1}{2e}\varepsilon^{\mu
\nu\rho\lambda}\gamma_{5}\gamma_{\nu}S_{\rho\lambda}. \label{Gravitino Eq Mot}%
\end{equation}
This on-shell structure follows easily from the open structure of the
superalgebra of simple supergravity. The nonclosure gauge functions $G_{\mu
\nu}$ and $G_{\mu}^{ab}$ can be straightforwardly computed from Eqs.
(\ref{SG symm}) and (\ref{SG ghost symm}) and are given by \cite{tahiri3}
\begin{align}
G_{\mu\nu}  &  =-\frac{1}{8}(\bar{c}\gamma_{a}c)\left(  \frac{1}{4}e_{\mu}%
^{b}e_{\nu b}\gamma^{a}-\frac{1}{2}ee^{\rho a}\varepsilon_{\mu\nu\rho\tau
}\gamma_{5}\gamma^{\tau}\right) \label{V_supergravity}\\
&  -\frac{1}{8}(\bar{c}\sigma_{ab}c)\left(  e_{\mu}^{a}e_{\nu}^{b}+\frac{1}%
{2}e_{\mu}^{c}e_{\nu c}\sigma^{ab}-\frac{1}{2}e\varepsilon_{\mu\nu\rho\tau
}e^{\rho a}e^{\tau b}\gamma_{5}\right)  ,\nonumber
\end{align}%
\begin{equation}
G_{\mu}^{ab}=-\frac{1}{8}(\bar{c}\gamma_{\mu}\sigma^{ab}\gamma_{5}c)\bar
{c}\gamma_{5}. \label{Z_supergravity}%
\end{equation}

This situation fits with the general framework presented in Sec. \ref{Sec2}.
In this context one has to point out that these nonclosure gauge functions are
related upon identities of type (\ref{BRST1}) and (\ref{BRST2}) (see Ref.
\cite{tahiri3}). Note that in the case of simple supergravity, the gauge
functions $G_{\mu\nu}$ and $G_{\mu}^{ab}$ can be related linearly by the
relation
\begin{equation}
G_{\mu}^{ab}=\frac{1}{2}e^{\rho a}e^{\nu b}\bar{c}\gamma_{\rho}G_{\nu\mu},
\end{equation}
which simply follows from the identity $G_{\mu\nu}\gamma_{\rho}c=\frac{1}%
{8}e\varepsilon_{\mu\nu\rho\tau}(\bar{c}\gamma^{\tau}c)\gamma_{5}c$. Thus, all
the nonclosure gauge functions and related equations can be written in terms
of $G_{\mu\nu}$; for example, Eq. (\ref{BRST1}) can be recast as%
\begin{align}
QG_{\mu\nu}  &  =c^{\rho}\partial_{\rho}G_{\mu\nu}+\partial_{\mu}c^{\rho
}G_{\rho\nu}+\partial_{\nu}c^{\rho}G_{\mu\rho}-\frac{1}{2}c^{ab}[\sigma
_{ab},G_{\mu\nu}]\nonumber\\
&  +\frac{1}{2}\bar{c}\gamma^{\rho}\psi_{\rho}G_{\mu\nu}+\frac{1}{2}\bar
{c}\gamma^{\rho}\psi_{\mu}G_{\rho\nu}+\frac{1}{2}\bar{c}\gamma^{\rho}\psi
_{\nu}G_{\mu\rho},
\end{align}
which gives, by the way, the BRST transformation of the gauge function
$G_{\mu\nu}$. This latter equation holds off shell (without relying on the
gravitino equation of motion) and thus indicates that no higher-order gauge
functions exist.

We are now able to apply the prescription presented in Sec. \ref{Sec3} in
order to introduce a suitable set of auxiliary fields allowing the
construction of an off-shell BRST operator. We first observe that the only
nonvanishing gauge functions are $G_{\mu\nu B}^{A}$ and $G_{\mu A}^{ab}$
(fermionic indices $A$ and $B$ are now exhibited). Thus, the off-shell BRST
operator defined by Eqs. (\ref{BRST75}) and (\ref{BRST76}) as well as the
definition of the extended invariant action (\ref{Sinv}) allow us to introduce
the set of 16 bosonic auxiliary fields $\eta^{\mu A}$ of ghost number -1. In
view of the prescription introduced in Sec. \ref{Sec3}, these auxiliary fields
are related to the gravitino degrees of freedom. No auxiliary fields related
to the vierbein (or graviton) may occur in this approach. The off-shell BRST
transformations read then%
\begin{align}
\Delta e_{\mu}^{a}  &  =Qe_{\mu}^{a},\nonumber\\
\Delta\psi_{\mu}^{A}  &  =Q\psi_{\mu}^{A}+G_{\mu\nu B}^{A}\eta^{\nu B},
\end{align}

\begin{align}
\Delta c^{\mu}  &  =Qc^{\mu},\nonumber\\
\Delta c^{ab}  &  =Qc^{ab}+G_{\mu A}^{ab}\eta^{\mu A},\nonumber\\
\Delta c  &  =Qc,
\end{align}

\begin{equation}
\Delta\eta^{\mu A}=c^{\nu}\partial_{\nu}\eta^{\mu A}+\partial^{\mu}c_{\nu}%
\eta^{\nu A}-\frac{1}{2}c^{ab}\sigma_{abB}^{A}\eta^{\mu B}-\frac{1}%
{2e}\varepsilon^{\mu\nu\rho\sigma}(\gamma_{5}\gamma_{\nu})_{\quad B}%
^{A}S_{\rho\sigma}^{B},
\end{equation}
ensuring that $\Delta^{2}=0$ on all fields. Moreover, the extended action%
\begin{equation}
S_{inv}=\frac{1}{2}ee_{a}^{\mu}e_{b}^{\nu}R_{\mu\nu}^{ab}-\frac{1}%
{4}\varepsilon^{\mu\nu\rho\sigma}\bar{\psi}_{\mu}\gamma_{5}\gamma_{\nu}%
S_{\rho\sigma}-\frac{e}{2}\bar{\eta}^{\mu}G_{\mu\nu}\eta^{\nu}
\label{Sinv_sugra}%
\end{equation}
is, up to a divergence term, invariant upon the action of $\Delta$. One can
also see that the fields$\ \eta^{\nu}$ are clearly nonpropagating fields since
their equations of motion are purely algebraic, i.e., $\delta S_{inv}%
/\delta\bar{\eta}^{\mu}=0$ leads to $G_{\mu\nu}\eta^{\nu}=0$.

\section{\bigskip Conclusion}

To conclude, in the present paper, we have given a prescription leading to the
construction of an off-shell BRST invariant quantum action for irreducible
open gauge theories \cite{batalin,Hen} described by a gauge algebra with
vanishing higher-order gauge functions. We first obtained an on-shell BRST
invariant quantum action containing four-ghost interaction terms typical for
open gauge theories as in supergravity theories. This follows from a gauge
-fixing action written as in Yang--Mills-type theories by modifying the
classical BRST operator . We then used a trick that permits us to introduce
auxiliary fields through the variation of the gauge-fixing fermion with
respect to the gauge fields, as turned out to be possible in BF theory
\cite{tahiri4}. Thus, we arrived at a closed BRST algebra together with an
off-shell invariant full quantum action in which, particularly, the invariant
extension of the classical action arose from the quartic ghost interaction terms.

As an application we show what our proposed construction gives in the case of
simple supergravity. The main result is that an off-shell realization can be
achieved upon the introduction of 16 auxiliary bosonic spinor variables of
ghost number -1. However, it is worthwhile to mention that the obtained set of
auxiliary fields are nonclassical fields, not only in the sense that they are
of nonzero ghost number but also because they do not balance the bosonic and
fermionic degrees of freedom off shell. Indeed, with 16 bosonic auxiliary
fields $\eta^{\mu A}$, one ends up at the off-shell level with 10 extra
bosonic degrees of freedom of ghost number -1, while, as usual, the bosonic
and fermionic degrees of freedom balance on shell. This can be compared with
the already-known off-shell formulations of simple supergravity which are,
namely, the minimal, new minimal, and non minimal (for a review, see Ref.
\cite{gates}). Even if these formulations differ in their auxiliary fields
structures, the total number of fermionic and bosonic degrees of freedom
balances at both on-shell and off-shell levels. One may note that it is
conceivable that the auxiliary fields introduced in this paper may be turned
into a set of \textquotedblleft classical\textquotedblright\ zero ghost
bosonic fields by taking advantage of a kind of twist redefinition
\cite{witten}; see also Ref. \cite{pestun} and references therein. This kind
of technique is (for instance) used in the context of superstring theory,
where auxiliary bosonic spinor variables can be treated as ghosts (or
antighosts) in the frame of the pure spinor approach by N. Berkovits; see,
e.g., Ref. \cite{matone} and references therein. If such an approach might
work, one should find some relations between the obtained auxiliary fields in
order to reduce them to the usual auxiliary fields sets of the known off-shell
formulations of simple supergravity. Such a construction will be analyzed in
details elsewhere.

\end{document}